\documentclass[11pt,fleqn]{article}

\usepackage{amsfonts}
\usepackage{amsmath}
\usepackage{amssymb}
\usepackage{mathtools}
\usepackage{mathrsfs}
\usepackage{enumerate}
\usepackage{bbm}
\usepackage{graphicx,epsfig,amsthm,color}
\usepackage{pstricks}
\usepackage{subfig}
\usepackage{graphicx}
\usepackage{units}

\usepackage{caption}
\usepackage{booktabs}
\usepackage{textcomp}
\usepackage{array}
\usepackage{cite}
\usepackage{cancel}

\date{\today}

\newcolumntype{z}[1]{>{\RaggedRight\hspace{0pt}}p{#1}}
\newcolumntype{w}[1]{>{\RaggedRight\hspace{0pt}}p{#1}}
\newcolumntype{v}[1]{>{\Centering\hspace{0pt}}p{#1}}

\def \la {\lambda}

\def \a {\alpha}

\def\be{\begin{equation}}
\def\ee{\end{equation}}
\def\bea{\begin{eqnarray}}
\def\eea{\end{eqnarray}}

\def\be{\begin{equation}}
\def\ee{\end{equation}}
\def\bea{\begin{eqnarray}}
\def\eea{\end{eqnarray}}

\def\erp2{{\rm e}^{2\rho}}
\def\erm2{{\rm e}^{-2\rho}}
\def\er4{{\rm e}^{4\rho}}

\def\be{\begin{equation}}
\def\ee{\end{equation}}
\def\bea{\begin{eqnarray}}
\def\eea{\end{eqnarray}}

\def\m0{m_{\nu_{0,i}}}

\def\T0{T_{\nu_0}}

\newcommand{\half}{\frac{1}{2}}

\newcommand{\beqa}{\begin{eqnarray}}
\newcommand{\eeqa}{\end{eqnarray}}
\newcommand{\bpr}{\begin{problem}}
\newcommand{\epr}{\end{problem}}
\newcommand{\bcent}{\begin{center}}
\newcommand{\ecent}{\end{center}}
\newcommand{\bfig}{\begin{figure}}
\newcommand{\efig}{\end{figure}}
\newcommand{\bpc}{\begin{picture}}
\newcommand{\epc}{\end{picture}}

\newcommand{\nnb}{\nonumber}
\newcommand{\reflef}{(\ref}

\renewcommand{\and}{A_{0}^{\nu ,D}(s)}

\newcommand{\bee}{\begin{equation}}

\def\beq{\begin{eqnarray}}
\def\eeq{\end{eqnarray}}

\newcommand{\Ga}{\Gamma}

\newcommand{\bright}{\begin{flushright}}
\newcommand{\eright}{\end{flushright}}
\newcommand{\bminip}{\begin{minipage}}
\newcommand{\eminip}{\end{minipage}}

\DeclareCaptionLabelSeparator{mysep}{\hspace{3pt}:\hspace{3pt}}
\DeclareCaptionLabelFormat{mypiccap}{Fig.\hspace{3pt}{#2}}
\DeclareCaptionLabelFormat{mytabcap}{Tab.\hspace{3pt}{#2}}

\begin{document}

\date{}
\title{
\vskip 2cm {\bf\huge Chameleonic dilaton and conformal transformations}\\[0.8cm]}

\author{
{\sc\normalsize Andrea Zanzi\footnote{Email: zanzi@th.physik.uni-bonn.de}\!\!}\\[1cm]
{\normalsize Via dei Pilastri 34, 50121 Firenze -  Italy}\\}
 \maketitle \thispagestyle{empty}
\begin{abstract}
{We recently proposed a chameleonic solution to the cosmological constant problem - Phys. Rev. D82 (2010) 044006. One of the results of that paper is a non-equivalence of different conformal frames at the quantum level. In this letter we further discuss our proposal focusing our attention on the conformal transformation. Moreover, we point out that a different choice of parameters is necessary in the model.  \\
}
\end{abstract}

\clearpage


\section{Introduction}

One of the main problems in modern Cosmology is the {\it cosmological constant} one \cite{Weinberg:1988cp} (for a review see
\cite{Nobbenhuis:2006yf}). In a recent paper \cite{Zanzi:2010rs} we solved this problem from the standpoint of string theory. The solution is obtained by mixing together some of the ideas currently known by the physics community to account for the cosmic accelerated expansion. Among them we mention: 1) a modification of GR at large distances (see for example \cite{Dvali:2000hr}); 2) backreaction effects \cite{Rasanen:2003fy, Kolb:2004am,
Kolb:2005da}; 3) a dynamic Dark Energy (DE) fluid. Let us start considering the element we
mentioned last. Scalar degrees of freedom are
a common feature in physics beyond the Standard Model (SM), for
example, they can be related to the presence of extra-dimensions. The Dark Energy could be the manifestation of an
ultralight scalar field rolling towards the minimum of its
potential \cite{Ratra:1987rm, Wetterich:1994bg, Zlatev:1998tr,
Carroll:1998zi}. Remarkably, there are reasons to maintain a
non-trivial coupling between the scalar field and matter, for
instance: a) to solve, at least partially, the coincidence
problem, a direct interaction between Dark Matter (DM) and DE has
been discussed \cite{Amendola:1999er, Amendola:2000uh,
TocchiniValentini:2001ty, Comelli:2003cv, Pietroni:2002ey,
Huey:2004qv, Amendola:2004ew, Gasperini:2003tf}; b) string theory
suggests the presence of scalar fields (dilaton and moduli)
coupled to matter (for an introduction see for example
\cite{Becker:2007zj, Gaillard:2007jr}). Consequently, a direct
interaction between matter and an ultralight scalar field can be
welcome. However, this could be phenomenologically dangerous:
violations of the equivalence principle (as far as the dilaton field is concerned the reader is referred to
\cite{Taylor:1988nw, Damour:1994ya, Damour:1994zq, Damour:2002nv, Damour:2002mi, Kaplan:2000hh, Damour:2010rp}), time dependence of couplings (for reviews
see \cite{Uzan:2010pm, Fischbach:1999bc}). 

One possible way-out is to consider "chameleon scalar fields"
\cite{Khoury:2003aq, Khoury:2003rn, Mota:2003tc},
namely scalar fields coupled to matter (including the baryonic
one) with gravitational (or even higher) strength and with a mass
dependent on the density of the environment. On cosmological
distances, where the densities are very small, the chameleons are
ultralight and they can roll on cosmological time scales. On the
Earth, on the contrary, the density is much higher and the field
is massive enough to satisfy all current experimental bounds on deviations from GR. 
In other words, the physical properties of this field
vary with the matter density of the environment and, therefore, it
has been called chameleon. The "chameleon mechanism" can be considered as a (local) 
stabilization mechanism which exploits the interaction matter-chameleon. Our solution to the cosmological constant problem discussed in \cite{Zanzi:2010rs} is obtained through these ideas: the solution is based on the chameleonic behaviour of the string dilaton\footnote{Many other stabilization
mechanisms have been studied for the string dilaton in the
literature. In particular, as far as heterotic string theory is
concerned, we can mention: the racetrack mechanism
\cite{Krasnikov:1987jj, Casas:1990qi}, the inclusion of
non-perturbative corrections to the Kaehler potential
\cite{Casas:1996zi, Binetruy:1996xja, Barreiro:1997rp}, the
inclusion of a downlifting sector \cite{Lowen:2008fm}, Casimir energy \cite{Zanzi:2006xr}.} \cite{Zanzi:2010rs}.

In the string frame (S-frame) of our model of reference \cite{Zanzi:2010rs}, the cosmological constant is very large and the dilaton is stabilized, while, after a conformal transformation to the Einstein frame (E-frame), the
dilaton is a chameleon and it is parametrizing the
amount of scale symmetry of the problem. Therefore, the E-frame cosmological constant is under control. This result points out a non-equivalence of different conformal frames at the quantum level (the cosmological constant is under control only in the E-frame). In the literature, scale invariance has
already been analyzed in connection to the cosmological constant problem (see for example \cite{Wetterich:2008sx, Wetterich:2009az, Wetterich:2008bf, Wetterich:2010kd} and references therein). In our
scenario \cite{Zanzi:2010rs}, the chameleonic behaviour of the field implies that
particle physics is the standard one only {\it locally}. All the
usual contributions to the vacuum energy (from supersymmetry
[SUSY] breaking, from axions, from electroweak symmetry
breaking...) are extremely large with respect to the meV-scale
only {\it locally}, while on cosmological distances (in the
E-frame) they are suppressed.

Typically, different conformal frames
are considered equivalent at the classical level and this result is
well-established in the literature (see for example
\cite{Catena:2006bd}). The main purpose of this paper is to further discuss the non-equivalence of different conformal frames at the quantum level and, in particular, to analyze carefully the conformal transformation. For further details on the
(non)-equivalence of different conformal frames the reader is
referred to \cite{Nojiri:2001pd, Fujii:2007qv, Alvarez:2001qj, Capozziello:2012cs} and
references therein.

As far as the organization of this letter is concerned, in section \ref{stsection} we briefly touch upon scalar-tensor theories of gravitation, in section \ref{CC} we write down the string frame action of our model and we touch upon the new choice of parameters. In section \ref{forum} we discuss the conformal transformation from the string frame to the Einstein frame in our model.

\setcounter{equation}{0}
\section{Scalar-Tensor theories of gravitation}
\label{stsection}

In this section we will briefly review some aspects of
Scalar-Tensor (ST) theories of gravitation following
\cite{Fujii:2003pa, Fujii:2004bg}.

\subsection{Jordan-Brans-Dicke models}

The Lagrangian of the original ST model by Jordan-Brans-Dicke (JBD) can be written in the form:
\begin{equation}
{\cal L}_{\rm JBD} = \sqrt{-g}\left(\half \xi \phi^2 R -\epsilon
\half g^{\mu\nu}\partial_\mu\phi \partial_\nu\phi +L_{\rm matter}
\right). \label{bsl1-4}
\end{equation}
$\xi$ is a dimensionless constant and $\epsilon=\pm1$ (in
particular $\epsilon=+1$ corresponds to a normal field having a
positive energy, in other words, not to a ghost). The convention
on the Minkowskian metric is (-,+,+,+). The first term on the
right-hand side is called "nonminimal coupling term" (NM), it is
unique to the ST theory and it replaces the Einstein-Hilbert term
(EH) in the standard theory:
\begin{equation}
{\cal L}_{\rm EH} = \sqrt{-g}\frac{1}{16\pi G}R. \label{bsl1-5}
\end{equation}
If we compare this last formula with the NM-term, we infer that in
this theory the gravitational constant $G$ is replaced by an
``effective gravitational constant" defined by
\begin{equation}
\frac{1}{8\pi G_{\rm eff}}= \xi \phi^2, \label{bsl1-6}
\end{equation}
which is spacetime-dependent through the scalar field $\phi(x)$.

We stress that Jordan admitted the scalar field to be included in
the matter Lagrangian $L_{\rm matter}$, whereas Brans and Dicke
(BD) assumed not. For this reason the name ``BD model'' seems
appropriate to the assumed {\it absence} of $\phi$ in $L_{\rm
matter}$.

\subsection{Conformal transformation}

\subsubsection{Scale transformation (Dilatation)}

Let us start with a global scale transformation in curved
spacetime, namely:

\begin{equation}
g_{\mu\nu}\rightarrow  g_{*\mu\nu}=\Omega^2
g_{\mu\nu},\quad\mbox{or}\quad g_{\mu\nu}=\Omega^{-2} g_{*\mu\nu},
\label{bsl1-45}
\end{equation}
where $\Omega$ is a constant, from which follows also
\begin{equation}
 g^{\mu\nu}=\Omega^{2} g^{*\mu\nu},\quad \mbox{and}\quad \sqrt{-g}=
 \Omega^{-4}\sqrt{-g_*}.
\label{bsl1-46}
\end{equation}

If we have only massless fields or particles, we have no way to
provide a fixed length scale, we then have a scale invariance or
dilatation symmetry. Had we considered a fundamental field or
particle having a nonzero mass $m$, the inverse $m^{-1}$ would
have provided a fixed length or time standard and the
above-mentioned invariance would have been consequently broken.

To implement this idea, let us introduce a real free {\it massive}
scalar field $\Phi$ (not to be confused with the dilaton), as a
representative of matter fields:
\begin{equation}
{\cal L}_{\rm matter}= \sqrt{-g}\left( -\half (\partial \Phi)^2
-\half m^2\Phi^2 \right), \quad (\partial \Phi)^2 \equiv
g^{\mu\nu}(\partial_\mu\Phi)(\partial_\nu\Phi). \label{bsl1-44}
\end{equation}
We then find
\begin{eqnarray}
{\cal L}_{\rm matter}&=&  \Omega^{-4}\sqrt{-g_*}\left(-\half
\Omega^2(\partial \Phi)^2 -\half m^2\Phi^2 \right) \nnb\\
&=& \sqrt{-g_*}\left(-\half (\partial_* \Phi_*)^2 -\half
\Omega^{-2}m^2\Phi_*^2 \right),\nnb\\
\quad \mbox{with}\quad \Phi_*=\Omega^{-1}\Phi. \label{bsl1-47}
\end{eqnarray}

Notice that we defined $\Phi_*$ primarily to leave the kinetic
term form invariant except for putting the $*$ symbol everywhere.
On the other hand, the mass term in the last equation breaks scale
invariance.

\subsubsection{Conformal transformation (Weyl rescaling)}

The global scale transformation in curved spacetime as discussed
above may be promoted to a {\it local} transformation by replacing
the constant parameter $\Omega$ by a local function $\Omega(x)$,
an arbitrary function of $x$.  This defines a conformal
transformation, or sometimes called Weyl rescaling:
\begin{equation}
 g_{\mu\nu}\rightarrow g_{*\mu\nu}=\Omega^{2}(x) g_{\mu\nu},\quad
 \mbox{or}\quad ds^2 \rightarrow ds^2_* = \Omega^{2}(x)ds^2.
\label{bsl1-51}
\end{equation}
According to the last equation, we are considering a local change
of units, not a coordinate transformation.  The condition for
invariance is somewhat more complicated than the global
predecessors.

Let us see how the ST theory is affected by the conformal
transformation.  We start with
\begin{equation}
\partial_{\mu}g_{\nu\lambda}=\partial_{\mu}\left( \Omega^{-2}g_{*\nu\lambda}
\right)=\Omega^{-2}\partial_{\mu}g_{*\nu\lambda} -2\Omega^{-3}\partial_{\mu}\Omega
g_{*\nu\lambda}=\Omega^{-2}\left( \partial_{\mu}g_{*\nu\lambda}
-2f_{\mu}g_{*\nu\lambda} \right), \label{bsl1-52}
\end{equation}
where $f=\ln\Omega, f_\mu =\partial_\mu f, f_*^\mu =
g_*^{\mu\nu}f_\nu$. We then compute
\begin{equation}
\Gamma^{\mu}_{\hspace{.3em}\nu\lambda}= \half g^{\mu\rho}\left(
    \partial_{\nu} g_{\rho\lambda}+\partial_{\lambda} g_{\rho\nu}
    -\partial_{\rho} g_{\nu\lambda} \right)
=\Gamma^{\mu}_{*\hspace{.1em}\nu\lambda}-\left(
f_{\nu}\delta^{\mu}_{\lambda}
  +f_{\lambda}\delta^{\mu}_{\nu}-f_{*}^{\mu}g_{*\nu\lambda}
\right), \label{bsl1-53}
\end{equation}
reaching finally
\begin{equation}
R=\Omega^2\left( R_{*}+6\Box_{*}f - 6
g_{*}^{\mu\nu}f_{\mu}f_{\nu}\right). \label{bsl1-54}
\end{equation}

Using this in the first term on the right-hand side of
\reflef{bsl1-4}) with $F(\phi) =\xi\phi^2$, we obtain
\begin{equation}
{\cal L}_1=\sqrt{-g} \half F(\phi) R =\sqrt{-g_{*}}\half
F(\phi)\Omega^{-2} \left( R_{*}+6\Box_{*}f - 6
g_{*}^{\mu\nu}f_{\mu}f_{\nu}\right). \label{bsl1-55}
\end{equation}
We may choose
\begin{equation}
F\Omega^{-2}=1, \label{bsl1-55a}
\end{equation}
so that the first term on the right-hand side goes to the standard
EH term.  We say that we have moved to the Einstein conformal
frame (E frame). We have
\begin{equation}
\Omega=F^{1/2},\quad\mbox{then } f= \ln\Omega,\hspace{1em}
f_{\mu}=\partial_\mu f=\frac{\partial_\mu\Omega}{\Omega}=\half
\frac{\partial_{\mu}F}{F} =\half\frac{F'}{F}\partial_{\mu}\phi,
\label{bsl1-56}
\end{equation}
where $F'\equiv dF/d\phi$. The second term on the right-hand side
of \reflef{bsl1-55}) then goes away by partial integration, while
the third term becomes
$-\sqrt{-g_*}(3/4)(F'/F)^2g_*^{\mu\nu}\partial_\mu\phi
\partial_\nu\phi$.  This term is added to the second term on the
right-hand side of \reflef{bsl1-4}) giving the kinetic term of
$\phi$:
\begin{equation}
-\half\sqrt{-g_{*}}\Delta
g^{\mu\nu}_{*}\partial_{\mu}\phi\partial_{\nu}\phi,
\quad\mbox{with}\quad \Delta= \frac{3}{2}\left( \frac{F'}{F}
\right)^2 +\epsilon\frac{1}{F}. \label{bsl1-57}
\end{equation}

If $\Delta >0$, we define a new field $\sigma$ by
\begin{equation}
\frac{d\sigma}{d\phi}= \sqrt{\Delta},\quad\mbox{hence}\quad
\sqrt{\Delta}\partial_\mu\phi =\frac{d\sigma}{d\phi}
\partial_\mu\phi = \partial_\mu\sigma, \label{bsl1-58}
\end{equation}
thus bringing \reflef{bsl1-57}) to a canonical form
$-(1/2)\sqrt{-g_{*}}g^{\mu\nu}_{*}\partial_{\mu}\sigma\partial_{\nu}\sigma$.
If $\Delta <0$, the opposite sign in the first expression of
\reflef{bsl1-57})  propagates to the sign of the preceding
expression, implying a ghost.

By using the explicit expression of $F(\phi)$ we find
\begin{equation}
\Delta = \left( 6+\epsilon\xi^{-1} \right)\phi^{-2} \equiv
\zeta^{-2}\phi^{-2}, \label{bsl1-59}
\end{equation}
which translates the condition $\Delta >0$ into $\zeta^2 >0$.  We
further obtain
\begin{equation}
\frac{d\sigma}{d\phi}=\zeta^{-1}\phi^{-1},\quad\mbox{hence}\quad
\zeta\sigma = \ln \left(\frac{\phi}{\phi_{0}}
\right),\quad\mbox{or}\quad
 \phi=\xi^{-1/2}e^{\zeta\sigma},
\label{bsl1-60}
\end{equation}
reaching also
\begin{equation}
\Omega =e^{\zeta\sigma}= \sqrt{\xi} \phi. \label{bsl1-60a}
\end{equation}
We finally obtain the lagrangian in the E frame:
\begin{equation}
{\cal L}_{\rm JBD}=\sqrt{-g_{*}}\left( \half R_{*} - \half
    g^{\mu\nu}_{*}\partial_{\mu}\sigma\partial_{\nu}\sigma
    +L_{\rm *{\rm matter}}  \right).
\label{bsl1-61}
\end{equation}

In the next section we will describe our model.

\setcounter{equation}{0}
\section{The model}
\label{CC}

In this section we will briefly summarize the lagrangian of our stringy solution to the
cosmological constant problem presented recently in \cite{Zanzi:2010rs} and we will point out that a different choice of parameters is necessary.

\subsection{The action}
\label{modello}

Our starting point is the string-frame, low-energy, gravi-dilaton
effective action, to lowest order in the $\a'$ expansion, but
including dilaton-dependent loop (and non-perturbative)
corrections, encoded in a few  ``form factors" $\psi(\phi)$,
$Z(\phi)$, $\alpha{(\phi)}$, $\dots$, and in an effective dilaton
potential $V(\phi)$ (obtained from non-perturbative effects). In
formulas (see for example \cite{Gasperini:2001pc} and references therein):
\bea S &=& -{M_s^{2}\over 2} \int d^{4}x \sqrt{-  g}~
\left[e^{-\psi(\phi)} R+ Z(\phi) \left(\nabla \phi\right)^2 +
{2\over M_s^{2}} V(\phi)\right]
\nonumber \\
&-& {1\over 16 \pi} \int d^{4}x {\sqrt{-  g}~  \over
\alpha{(\phi)}} F_{{\mu\nu}}^{2} + \Ga_{m} (\phi,  g, \rm{matter})
\label{3} \eea Here $M_s^{-1} = \la_s$ is the fundamental
string-length parameter and $F_{\mu\nu}$ is the gauge field
strength of some fundamental grand unified theory (GUT) group ($\a(\phi)$ is the
corresponding gauge coupling). We imagine having already
compactified the extra dimensions and having frozen the corresponding
moduli at the string scale.

Since the form factors are {\it unknown} in the strong coupling
regime, we are free to {\it assume} that the structure of these
functions in the strong coupling region implies an S-frame
Lagrangian composed of two different parts: 1) a scale-invariant
Lagrangian ${\cal L}_{SI}$. This part of our lagrangian has
already been discussed in the literature by Fujii in references
\cite{Fujii:2002sb, Fujii:2003pa}; 2) a Lagrangian which
explicitly violates scale-invariance ${\cal L}_{SB}$.

In formulas we write:

\beq {\cal L}={\cal L}_{SI} + {\cal L}_{SB}, \label{Ltotale}\eeq where the
scale-invariant Lagrangian is given by:

\begin{equation}
{\cal L}_{\rm SI}=\sqrt{-g}\left( \half \xi\phi^2 R -
    \half\epsilon g^{\mu\nu}\partial_{\mu}\phi\partial_{\nu}\phi -\half g^{\mu\nu}\partial_\mu\Phi \partial_\nu\Phi
    - \frac{1}{4} f \phi^2\Phi^2 - \frac{\lambda_{\Phi}}{4!} \Phi^4
    \right).
\label{bsl1-96}
\end{equation}
$\Phi$ is a scalar field representative of matter fields,
$\epsilon=-1$, $\left( 6+\epsilon\xi^{-1} \right)\equiv
\zeta^{-2}\simeq 1$, $f<0$ and $\lambda_{\Phi}>0$.
One may write also terms like $\phi^3 \Phi$, $\phi \Phi^3$ and
$\phi^4$ which are multiplied by dimensionless couplings. However
we will not include these terms in the lagrangian. The symmetry breaking Lagrangian
${\cal L_{SB}}$ is supposed to contain scale-non-invariant terms,
in particular, a stabilizing (stringy) potential for $\phi$ in the
S-frame. For this reason we write: \beq {\cal L}_{\rm
SB}=-\sqrt{-g} (a \phi^2 + b + c \frac{1}{\phi^2}). \label{SB}
\eeq

Happily, it is possible to satisfy the field equations with
constant values of the fields $\phi$ and $\Phi$ through a proper
choice (but not fine-tuned) values of the parameters
$a, b, c$, maintaining $f<0$ and $\lambda_{\Phi}>0$. We made sure
that $g_s>1$ can be recovered in the equilibrium configuration and
that, consequently, the solution is consistent with the
non-perturbative action that we considered as a starting point.

Here is a possible choice of parameters (in string units):
$f=-2/45$, $\lambda_{\Phi}=0.3$, $a=1$, $b=-\frac{13}{72}$, $c=1/108$, $\zeta=5$. In the equilibrium configuration we have
$\phi_0=\frac{1}{2}$ and $\Phi_0=\frac{1}{3}$. In the next paragraph we calculate the S-frame field equations and we discuss the constraints which brought us to our choice of parameters.

\subsection{S-frame metric and field equations}

As far as the dilaton is concerned, we calculate the variation of the Lagrangian and we write:

\bea
\xi \phi R + \epsilon \Box \phi -\frac{f}{2} \phi \Phi^2 -2a\phi +\frac{2c}{\phi^3}=0.
\label{prima}
\eea
Multiplying by $\phi$ we have
\bea
\xi \phi^2 R+\epsilon \phi \Box \phi-\frac{f}{2} \phi^2 \Phi^2-2a\phi^2+\frac{2c}{\phi^2}=0.
\label{seconda}
\eea

If we define $\varphi=\frac{1}{2} \xi \phi^2$, ''Einstein'' equations become

\bea
2 \varphi G_{\mu\nu}=T_{\mu\nu}-2(g_{\mu\nu}\Box-\nabla_{\mu}\nabla_{\nu})\varphi,
\label{terza}
\eea
where $T_{\mu\nu}$ is the total energy momentum tensor. Taking the trace, we can write
\bea
-2 \varphi R=-(\partial\Phi)^2-f\phi^2\Phi^2-\frac{\lambda}{3!}\Phi^4-4a\phi^2-4b-\frac{4c}{\phi^2}-\epsilon(\partial\phi)^2 -6\Box\varphi.
\label{quarta}
\eea
If we add together \ref{seconda} with \ref{quarta}, we obtain

\bea
\epsilon \phi \Box \phi +\frac{f}{2} \phi^2 \Phi^2 +2a \phi^2 +\frac{6c}{\phi^2}=-\epsilon (\partial \phi)^2 -6 \Box \varphi-(\partial \Phi)^2 -\frac{\lambda}{3!} \Phi^4-4b
\label{quinta}
\eea
 and remembering that $\epsilon[\phi \Box \phi+ (\partial \phi)^2]=\frac{\epsilon}{2} \Box \phi^2$ we obtain the final dilatonic equation as
 
\bea
(6+\epsilon \xi^{-1}) \Box \varphi + (\partial \Phi)^2+ \frac{f}{2} \phi^2 \Phi^2 + \frac{\lambda}{3!} \Phi^4+ 2 a \phi^2+ 4 b +\frac{6 c}{\phi^2}=0.
\label{finaledilatone}
\eea

As far as the matter field is concerned we use \cite{Fujii:2003pa}

\bea
\Box \Phi -\frac{f}{2} \phi^2 \Phi-\frac{\lambda}{6}\Phi^3=0.
\label{matterf}
\eea
 
The parameters are chosen exploiting the following constraints, namely:
\begin{itemize}
\item The stationarity condition for matter fields
\bea
\Phi^2=-\frac{3f}{\lambda} \phi^2.
\label{station}
\eea
\item A stationarity condition for the dilaton
\bea
\frac{f}{2}\phi^2\Phi^2+\frac{\lambda}{6} \Phi^4 +2 a \phi^2+ 4 b+\frac{6 c}{\phi^2}=0.
\eea
\end{itemize}
One more remark is necessary. From formula \ref{quarta} we see that the 4-dimensional curvature in the S-frame is {\it constant}. With our choice of parameters we find a positive curvature, $R\simeq10.1$ (in dimensionless units). Therefore we choose the de Sitter metric as our S-frame metric.

\setcounter{equation}{0}
\section{Discussion: the conformal transformation and non-equivalent frames}
\label{forum}

Remarkably, even if we stabilize the dilaton in
the S-frame, the conformal transformation to the E-frame is
non-trivial. This point needs to be further elaborated. First of all, even if we considered a {\it classical} field theory, a dependence of the dynamical behaviour of the fields on the choice of the conformal frame would be possible in certain cases. Therefore, let us start considering a {\it classical} field theory with a scalar field (that we call dilaton) whose dynamical behaviour is governed by a lagrangian which is formally equivalent to our S-frame lagrangian. Let us
consider a stabilizing potential $V(\sigma)$ for the dilaton in
the S-frame and let us call $\sigma_0$ the value of the dilaton in
the minimum of the potential. This constant value can be called the (vacuum) expectation value of the field and it is {\it classical}. When we perform the conformal
transformation, we write $\phi=\xi^{-1/2} M_p e^{\zeta \sigma}$, where $M_p$ can be simply considered a mass parameter of this classical theory. The minimum of the potential will be
multiplied by the conformal factor $e^{-4 \zeta \sigma_0}$ (which
is constant). A different point of the potential, for example
$V(\sigma_1)$, will be multiplied by a different constant
conformal factor (i.e. $e^{-4 \zeta \sigma_1}$). Consequently, the
potential will be multiplied by a non-constant function
of the dilaton, namely, a non-trivial conformal factor given by $\xi^{-2} \phi^{-4}=e^{-4 \zeta \sigma}$.
The potential \ref{SB} will be mapped by the conformal transformation into an E-frame potential given by
\beq V_{SB}= e^{-4\zeta\sigma}[a\phi^2+b+\frac{c}{\phi^2} ].
\label{totdil}
\eeq 

Summarizing, in this classical field theory example, a constant $M_p$-parameter guarantees a one-to-one map between the classical vacuum in the first frame and one single classical vacuum in the second frame. Moreover $\phi$ and $\sigma$ are linked together: no matter which conformal frame we choose, we can identify the dilaton equivalently with $\phi$ or $\sigma$, but in general a stabilized dilaton (call it $\phi$ or $\sigma$) in one frame does not correspond to a stabilized dilaton (call it $\phi$ or $\sigma$) in a different frame. Let us now discuss the {\it quantum} field theory case and let us come back to our model of reference \cite{Zanzi:2010rs}. When we perform the conformal transformation, we can formally consider the Planck mass as a constant parameter (which is fixed to be equal to one) and, simultaneously, non-constant values of the dilatons $\phi$ and $\sigma$ (linked to each other). This constant Planck mass is {\it unrenormalized} and it is the relevant one when we deal with a {\it function} of the dilaton (and not with its expectation value). On the contrary, if our intention is to create a connection with the {\it physical renormalized} Planck mass, we must give an expectation value to the dilaton field. Since the S-frame dilaton is constant, $\phi=\phi_0=1/2$, we infer that right after the conformal transformation (i.e. after the first quantization step - see also \cite{Zanzi:2010rs}) the physical Planck mass becomes an {\it exponentially} decreasing function of the expectation value of $\sigma$. The next question is: what about the dynamical behaviour of $\sigma$ (i.e. what about its potential in the E-frame)? The physical renormalized $V_{SB}$ is certainly run-away towards large $\sigma$ with our choice of parameters and fields (notice that the square bracket in \ref{totdil} is positive and constant). The renormalized Planck mass is a decreasing function of $\sigma$, therefore, the Einstein-Hilbert term is compatible with the restoration of scale invariance for large $\sigma$ discussed in \cite{Zanzi:2010rs} which is the crucial element to obtain a chameleonic behaviour of the dilaton in the E-frame. This exponentially decreasing Planck mass is renormalized and its non-constant nature implies that a single {\it quantum} vacuum in the S-frame corresponds to an {\it infinite number} of different {\it quantum} vacua in the E-frame. The concept of vacuum is different when we move from a classical to a quantum field theory and the renormalization of the Planck mass breaks the one-to-one link between vacua discussed in the classical case. As already mentioned in \cite{Zanzi:2010rs}, locally we get a large contribution to the vacuum energy from matter fields and this contribution is planckian. We can consider many small local bubbles with a large vacuum energy and when we average these contributions on very large (i.e. cosmological) distances we obtain a large unrenormalized contribution to the vacuum energy (and this fact is related to a constant unrenormalized Planck mass). On the contrary, the corresponding renormalized contribution is obtained by giving an expectation value to the dilaton, it is exponentially suppressed for large values of $\sigma$ and it is fully compatible with the restoration of scale invariance for large sigma, namely:
\bea M_p \propto  e^{-\zeta \sigma}.\eea

We warn the reader that in this letter, as far as the quantization of the theory is concerned, we considered only what we called step 1. On the other hand, it seems worthwhile to point out that our claim of reference \cite{Zanzi:2010rs} regarding a (almost) negligible Einstein-Hilbert term on cosmological distances takes into account also the next quantization steps. These issues will be further discussed in a future work.

\subsection*{Acknowledgements}

I warmly thank Gianguido Dall'Agata, Antonio Masiero, Marco Matone, Massimo Pietroni and Roberto Volpato for useful conversations.


\providecommand{\href}[2]{#2}\begingroup\raggedright\endgroup

\end{document}